\documentclass[twocolumn, prl]{revtex4}
\usepackage{graphicx}

\begin{document}
\title{Spectral analysis of short time signals}
\author{Zbyszek P. Karkuszewski}

\affiliation{
Institute of Physics, Jagiellonian University, Cracow, Poland\\
and\\
Los Alamos National Laboratory, Los Alamos, NM 87545.
}

\date{\today}

\begin{abstract}
The very old problem of extracting frequencies from time signals is addressed
  in the case of signals that are very short as compared to their 
  intrinsic time scales.
The solution of the problem is not only important to the classic signal processing but also helps to disqualify several common formulations of the quantum
mechanical time-energy uncertainty principle.
\\ \\
LAUR-04-5290
\end{abstract}

\maketitle

\section{Introduction}
The goal of many scientific efforts is to predict future evolution of
physical systems on the basis of their known past behavior.
One example of the very limited success of such an activity is
weather forecasts. Even if the history of all important parameters 
like temperature, pressure, humidity, wind velocity, etc. is known for many 
years back at almost every point on Earth, the reasonably accurate prediction 
of the coming weather conditions can be made only for several days.
One may argue that Earth's atmosphere is especially tough system to consider
due to its intrinsic instabilities: Even a tiny perturbation of air in one 
place can lead to huge changes of weather on a distant continent. In this
work we will not be able to deal with such instable systems either.

The other extreme is represented by very stable systems such as 
celestial objects. Centuries-long observations of the Moon and the Sun allowed 
ancient astronomers to predict accurately, Moon's phases, risings, settings 
and eclipses for coming millennia without any knowledge of
gravitational forces or Kepler's laws. Such precision was possible 
because the 
observations of the system had been made over much longer period than the 
system's characteristic time scales: days, (sidereal) months and years. 

Would the same quality predictions be possible if we observed the Moon 
just for fifteen minutes, i.e. for time much shorter than the shortest 
characteristic timescale?

The situation is even more interesting in quantum world where according
to some formulations of the time-energy uncertainty principle it would be
fundamentally impossible to accurately predict evolution of a quantum system
that has been observed only for a very short period of time.

In this work we show a practical way of achieving exact predictions of a future
based on a very short history of a system. Our predictions will be limited
only to the quantities evolution of which can be well described by finite 
Fourier series. This restriction is crucial. Still, there are many important
quantities that fall into this category.

Before introducing a method designed to perform such a task let us restate the
problem in a more formal way.
Suppose that a continuous quantity (a signal) $c(t)$ is given only in a finite 
time interval $t\in [0,T]$ and that it can be expressed in the following way 
\begin{equation}
c(t) = \sum_{k=1}^K d_k e^{-i\omega_k t},
\label{sigc}
\end{equation}
where $K$ is an integer, $d_k$ is a real positive \cite{ssp} amplitude 
and $\omega_k$ is a real (characteristic) frequency. The quantity $c(t)$ is, 
for simplicity, a complex function, but it can be made real by appropriate 
addition of terms with opposite sign frequencies $\omega_{k'}=-\omega_k$.

Our goal is to find unknown amplitudes $d_k$ and frequencies $\omega_k$ of 
given $c(t)$ in the case when length $T$ of the time interval is much smaller
than the smallest characteristic timescale $T\ll 1/\omega_{max}$. The method
outlined bellow does not require the prior knowledge of the number of the
Fourier components $K$ in (\ref{sigc}).

Mathematically, one can see that $c(t)$ is an analytic function of
time $t$ and, as such, can be uniquely extended beyond the interval $[0,T]$.
This means that, in principle, even for tiny $T$ it is possible to get all
amplitudes and frequencies from (\ref{sigc}) to any desired precision.  
Unfortunately it is difficult to solve this nonlinear problem analytically
and numerical methods cannot handle continuous signals due to infinite number
of data points.  One way of getting around this problem is to take only finite 
number of points at the cost of loss of the uniqueness of the extension.
From now on we will assume that the signal $c(t)$ is
known only at $N+1$ equidistant time points $t_n=n\delta t$ for $n=0,...,N$
and $t_N=T$. (\ref{sigc}) can be rewritten as a set of $N+1$ equations
\begin{equation}
\sum_{k=1}^K d_k e^{-i\omega_k t_n} = c_n,
\label{sigd}
\end{equation}
where $c_n\equiv c(n\delta t)$. This set has $2K$ real unknowns 
($K$ amplitudes and $K$ frequencies) so the number of complex equations 
$N+1$ has to be equal or greater than $K$. This condition would have opened the possibility
of existence of a unique solution if the equations were linear in $\omega_k$ 
and $d_k$. It is not the case here. There will always be infinite number of solutions
to (\ref{sigd}) if the set is self-consistent and no solutions otherwise.
If one has an additional information about the range of frequencies $\omega_k$
in the problem (for instance Moon trajectory on the sky should not involve 
frequencies higher than 1/Hour) then the solution will be unique for small 
enough $\delta t$.

Similar problems, but for large $T$, are usually treated by a
discrete Fourier transform method (DFT).
That method {\it assumes} a grid of $N$ equidistant frequencies and solves 
(\ref{sigd}) only for $d_k$ as a linear set of equations. The spacing
of the assumed frequencies is proportional to $1/T$ and thus the method
is useless for estimating values of $c$ outside very short interval $T$.

The more challenging task of solving (\ref{sigd}) for both
amplitudes and frequencies is performed by, so called, harmonic inversion 
method.

\section{Harmonic inversion}
The historical roots of the method can be traced back to the two 
centuries old work by Gaspard Riche (Baron de Prony) \cite{Prony}.
A numerical implementation of the original approach can be found 
in \cite{Marple}. Here we present yet another derivation of the algorithm.

The key idea behind the harmonic inversion method is to replace the 
nonlinear problem (\ref{sigd}) with an eigenvalue problem for some
fictitious operator $\hat U$.

It is not necessary but very convenient to use basic formalism of quantum 
mechanics to introduce mathematical structure of the method. We will also 
benefit from this formalism when we turn our attention to quantum time-energy 
uncertainty principles.

Suppose that an evolution of a normalized quantum state $|\Phi_0\rangle$ 
is generated by an unitary operator $\hat U(\delta t)$ and 
$|\Phi_n\rangle \equiv \hat U^n(\delta t)|\Phi_0\rangle$. 
For every signal $c$ in (\ref{sigd}) there exists such an 
evolution operator $\hat U$ that the signal can be presented as 
the following autocorrelation function
\begin{equation}
c_n = \langle \Phi_0| \Phi_n\rangle.
\label{auto}
\end{equation}
It is enough to find eigenvalues $u_k=\exp(-i\omega_k\delta t)$ of the operator 
$U(\delta t)$ to find all characteristic frequencies. 
The matrix elements of $\hat U$ in the basis of states
$|\Phi_n \rangle$ can be expressed in terms of $c_n$ alone 
\begin{equation}
U_{ij}\equiv \langle \Phi_i | U| \Phi_j\rangle = 
\langle \Phi_0| U^{j-i+1} | \Phi_0\rangle = c_{j-i+1} 
\label{me}
\end{equation}
where $i,j=0,..., N-1$. 
The negative indices of $c$ in the equation above introduce no complication 
since $c_{-n}=c^*_n$. To obtain $K$ eigenvalues the dimension $N$ of the matrix 
$U$ must be equal or greater than $K$. This means that the number of
complex signal points $c_n$ required by the method exceeds the half of the 
number of unknowns. Vectors $|\Phi_n\rangle$ are not orthogonal,
thus the eigenequation for matrix $U$ takes the form
\begin{equation}
U |u_k\rangle = u_k S |u_k\rangle,
\label{gep}
\end{equation}
where $S$ is a matrix of scalar products $S_{ij}\equiv\langle
\Phi_i|\Phi_j\rangle=c_{j-i}$ with $i,j=0,...,N-1$. The rank of the matrix
$S$ gives the number of Fourier components of the signal $K$ for $N\ge K$.

The harmonic inversion method consists of two stages. First, one numerically 
solves the generalized eigenvalue problem (\ref{gep}), in order to get all 
frequencies. Second, when all frequencies in (\ref{sigd}) are known, a 
linear set of equations for the amplitudes $d_k$ is solved.

This brilliant algorithm has been developed and used by physicists and 
chemists for several years now, \cite{Neuhauser,Taylor1,Taylor2}. The 
only problem is that it fails when applied to short (small $T$) 
signals, \cite{Taylor2}. This limitation has been phrased in a form of 
Fourier-like uncertainty relation stating that the local density of 
frequencies that can be resolved by harmonic inversion must be smaller than the length $T$ of the time span of the signal \cite{Taylor2}.

Here we claim that the applicability of the method is not limited by the length
$T$ of the time interval but rather by noise affecting the signal $c(t)$.

In short, solving (\ref{gep}) requires calculating an inverse of the Hermitian 
scalar product matrix $S$. $S$ has $K$ positive and $N-K$ zero eigenvalues.
There are algebraic techniques to cope with singular matrices, see \cite{Golub}.
However, the smallest positive eigenvalue $\lambda_{min}$ becomes very small
for small $T$ or large $K$.
\begin{equation}
\frac{\lambda_{min}}{KN} \approx [f(\omega_k)T  \Omega]^{2(K-1)}
\label{mineig}
\end{equation}
where $f(\omega_k)$ is a function of order of unity,
and $\Omega$ stands for the greatest frequency magnitude present in the signal.
The formula above is valid only for short signals, $\Omega T \ll 1$, and the
expression in rectangular brackets is less than 1. 

Even small addition of noise to the signal may result in such a perturbation
of $\lambda_{min}$ that it will be impossible to distinguish it from, also 
perturbed, zero eigenvalues of the matrix $S$. When that happens harmonic 
inversion fails. The optimistic approach to (\ref{mineig}) notices that
$\lambda_{min}$ increases very fast with increasing $T$.

Assuming that the signal $c(t)$ is corrupted by noise $\eta(t), 
\quad\eta(t)\in [-\eta_{max}, \eta_{max}]$, the new 
signal $\tilde c_n = c_n + \eta_n$ has to be used to build matrices in 
(\ref{gep}). Harmonic inversion will extract all frequencies if
\begin{equation}
\lambda_{min} \ge 4N\eta_{max}
\label{nc}
\end{equation}
which is a necessary condition assuring that all $K$ positive eigenvalues 
of $S$ can be found.
Presence of noise leads to a set of perturbed frequencies 
$\tilde \omega_k$, which differ from $\omega_k$ 
\begin{equation}
|\tilde \omega_k - \omega_k|T \le \frac{2KN^2}{\lambda_{min}}\eta_{max}.
\label{ina}
\end{equation}
Notice that this is a "certainty" rather than uncertainty relation. 
The accuracy of frequencies found with the harmonic inversion can be made 
as high as needed by reducing the amplitude of noise $\eta_{max}$. 
This is the central result of this work.

\begin{figure}[htb]
\includegraphics*[width=8.6cm]{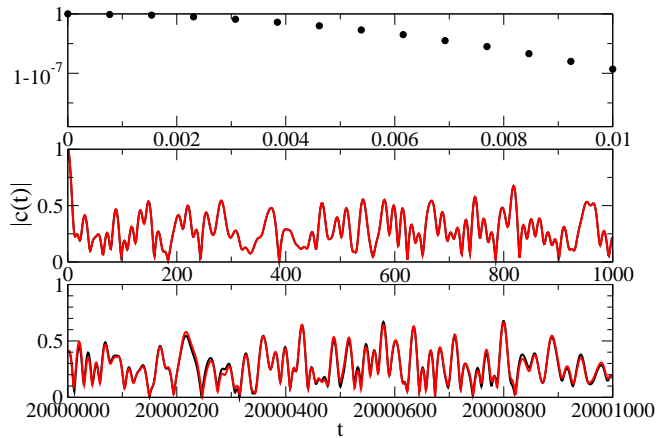}
\caption{As an example, the harmonic inversion method was applied
to a signal with $K=10$ frequencies drawn from the interval $(0.5,1.0)$,
sampled at $N+1=14$ points with $T=0.01$ (upper plot). 
85 digits precision, i.e. $\eta_{max}=10^{-84}$, was used.
Based on the points of the upper plot the signal is reconstructed (red line)
with the help of harmonic inversion.
Exact (black line) and reconstructed (red line) signals are indistinguishable
(middle plot).
In fact, the two lines start to differ by 1\% for $t>1,000,000$.
The discrepancies, that are due to roundoff errors (noise) of the initial 
points, are visible in the bottom plot.
In the example above $\lambda_{min}=4.07\times 10^{-78}$ and this justifies
the used 85 digits precision.
As a curiosity, the DFT algorithm applied to this signal would give just one 
frequency $\omega=0$, so that the prediction would be a horizontal
line at $|c(t)|=1$.}
\label{Fig1}
\end{figure}

In numerical studies where the method failed for small $T$ the role of noise was
played by roundoff errors. Figure \ref{Fig1} shows an exemplary application
of the harmonic inversion to a very short signal. The values of 
parameters in the example were deliberately chosen to expose the importance 
of the precision of the signal.

Coming back to the problem of determining the future positions of the Moon in 
the sky just from a short observation we see that it is possible under
one condition: The observation has to be very accurate. The major obstacle
in achieving required accuracy would be refraction of the incoming moonbeams 
in the Earth's atmosphere, which results in a significant shift of the
apparent Moon's position with respect to the actual position \cite{Fantz}.

The method described above can be applied to any signal of classical
or quantum origin as long as it has the form of (\ref{sigd}).
When applied to quantum systems, it gives important insight to the problem
of validity of several formulations of the time-energy uncertainty relations.

\section{Quantum uncertainty relations}
Uncertainty principles play a central role in quantum mechanics.
They impose constraints on the states allowed by the theory. For example,
no quantum state can yield a product of momentum and position standard 
deviations smaller than $\hbar/2$ i.e. $\Delta p\Delta x\ge \hbar/2$,
where momentum $p$ and position $x$ are a pair of canonically conjugate 
operators $[x,p]=i\hbar$. This also means that if we prepare a large number of
quantum systems, all in the same state, and perform an {\it exact} 
measurement of position on half of them and an {\it exact} measurement of 
momentum on the other half then the spread of measured values must satisfy the 
uncertainty principle above.

Similar in form is the relation often provided for time $t$ and energy $E$
\begin{equation}
\Delta E\Delta t \ge \hbar/2,
\label{dedt}
\end{equation}
which is interpreted in various ways in literature. Unlike energy, 
time is just a parameter in quantum mechanics and the analogy between 
(\ref{dedt}) and other uncertainty relations cannot be carried too far.
One of the formulations of the time-energy uncertainty relation is presented
in textbooks on quantum mechanics \cite{Schiff,Messiah} as follows:
The measurement of energy of a quantum system performed over time $\Delta t$ 
inevitably results in inaccuracy $\Delta E$, so that (\ref{dedt}) is
satisfied. For more examples of absurd interpretations of the uncertainty
see \cite{Asher}.

It has been pointed out by Aharonov and Bohm \cite{Aharonov1} that the above
interpretation is wrong and one can measure accurate energy
of a quantum system in as short time interval as one pleases.
With the use of harmonic inversion we can provide a simple argument supporting
that claim. Moreover, the argument outlined bellow is much simpler and more
general than the one used in \cite{Aharonov1}.

Suppose that the state of an isolated quantum system spans over a finite
number $K$ of its energy eigenstates $|k\rangle$, so that at any time $t$
it assumes the form
\begin{equation}
|\Phi(t)\rangle = \sum_{k=1}^K a_k e^{-i \omega_k t} |k\rangle,
\end{equation}
where $a_k$ is a complex number and eigenenergy $E_k=\hbar\omega_k$.

We will show that, in principle, not only an expectation value of the 
energy but all energies $E_k$ that contribute to the evolution of the system
can be measured exactly no matter how short the measurement time is.

First, it has been experimentally proved that one can actually measure
a wave function $\Phi(x)$ in position representation. The technique used in
the measurement is known as quantum tomography. The relevant theory and 
applications are reviewed in \cite{Raymer}.

Second, being able to measure $\Phi(x)$ at different times implies that the time
autocorrelation function itself $c(t)=\langle\Phi(0)|\Phi(t)\rangle$ can be measured.
Moreover, accuracy of the estimates of $c(t)$ improves systematically with 
the increasing number of copies of the system on which such measurements are
performed. 

The third and the last step is to use harmonic inversion described above
to extract all eigenenergies and respective amplitudes from $c(t)$.

In short, the accuracy of obtained eigenenergies $E_k$ is only limited by
the precision of the measured autocorrelation function $c(t)$ and this precision
can be, in principle, as high as one needs.

\section{Unbreakable relation}
Many formulations of the time-energy uncertainty principle were invented using
intuition or dimensional analysis. And, as in the case discussed above, 
they are of limited applicability or simply wrong. There is, however,
one formulation that is rigorously derived from the quantum theory
\cite{Mandelshtam}.  

The Heisenberg uncertainty relations are manifestations of the following
theorem:
If $\hat A$ and $\hat B$ are two self-adjoint operators and a state
$|\Psi\rangle$ belongs simultaneously to the domains of $\hat A$, $\hat B$,
$\hat A\hat B$, $\hat B\hat A$, $\hat A^2$ and $\hat B^2$, then
\begin{equation}
\Delta A\Delta B \ge \frac{1}{2} |\langle [\hat A, \hat B]\rangle|,
\label{Hut}
\end{equation}
where $(\Delta A)^2\equiv \langle \hat A^2\rangle - \langle \hat A\rangle^2$
and $\langle ...\rangle\equiv \langle \Psi| ...|\Psi\rangle $. 
The uncertainty above is an intrinsic feature of the state $|\Psi\rangle$ and
has nothing to do with measurement inaccuracies.

This theorem applied to the position and momentum operators yields familiar
relation for standard deviations of position and momentum. In the case of time
and energy, however, (\ref{Hut}) results in $0\ge 0$ since time enters the
Schr\"odinger equation as a parameter rather than an operator. On the other
hand if $\hat A$ in (\ref{Hut}) is replaced with a Hamiltonian $\hat H$ and 
$|\Phi\rangle$ is not a stationary state, then using Heisenberg equation of
motion for an incompatible operator $\hat B$, 
$\mbox{d}\langle\hat B\rangle/\mbox{d}t = i/\hbar \langle [\hat H, \hat
B]\rangle$, one arrives at
\begin{equation}
\Delta H\frac{\Delta B}{\left | \frac{\mbox{d}\langle \hat B\rangle}
{\mbox{d}t}\right |} \ge \frac{\hbar}{2}
\end{equation}
the uncertainty relation of energy and something that has dimension of time --
the lifetime of the state $|\Psi\rangle$ with respect to the observable $B$.
This uncertainty relation cannot be broken or circumvented, it holds as
long as quantum mechanics is valid.

There is also a time-energy relation introduced for the case where only one
copy of a quantum system is used \cite{Aharonov2}.

\section{Summary}

In this work we show a practical way of extracting accurate frequencies and 
amplitudes from a signal that is available only over very short period of time.
The price is that the signal itself must be known to a very high
precision. The required precision is not achievable in real world
experiments where observed signals are both short and involve many frequencies.
The situation is somewhat better in numerical simulations where one has
more control over generated data. 
The role of this letter is to identify the source of the difficulty with
a spectral decomposition of short signals and present a tool to perform
such a decomposition when possible.

The method equips us also with a powerful argument against some
interpretations of time-energy uncertainty relations in quantum mechanics.

Harmonic inversion can also be applied to signals with continuous spectra.
In this case it provides lowest moments of the relevant frequency
distribution.

Harmonic inversion is superior to DFT even when long time signals are 
considered. Readers interested in testing the method on long signals and 
where complex frequencies might be involved should read \cite{Taylor2}.

\section{Acknowledgments}
I am grateful to Jacek Dziarmaga, Krzysztof Sacha, Jakub Zakrzewski and
George Zweig for many stimulating discussions.
Work supported by KBN grant 5 P03B 088 21.


\begin{thebibliography}{}
\bibitem{ssp} Negative amplitudes
would only require replacing an ordinary scalar product with a symmetric 
scalar product in this Letter, see \cite{Taylor2} for details.
\end{thebibliography}
\end{document}